# On the Origins of Near-Surface Stresses in Silicon around Cu-filled and CNT-filled Through Silicon Vias

Ye Zhu, Kaushik Ghosh, Hong Yu Li, Yiheng Lin, Chuan Seng Tan and Guangrui (Maggie) Xia

**Abstract**

Micro-Raman spectroscopy was employed to study the near-surface stress distributions and origins in Si around through silicon vias (TSVs) at both room temperature and elevated temperatures for Cu-filled and CNT-filled TSV samples. From the observations, we proved that the stresses near TSVs are mainly from two sources: 1) pre-existing stress before via filling, and 2) coefficients of thermal expansion (CTE) mismatch-induced stress. CTE-mismatch-induced stress is shown to dominate the compressive regime of the near-surface stress distribution around Cu-filled TSV structures, while pre-existing stress dominates the full range of the stress distribution in the CNT-filled TSV structures. Once the pre-existing stress is minimized, the total stress around CNT-filled TSVs can be minimized accordingly. Therefore, compared to Cu-filled TSVs, CNT-filled TSVs hold the potential to circumvent the hassle of stress-aware circuit layout and to solve the stress-related reliability issues.

## 1. Introduction

Through Silicon Via (TSV) technology has gained more and more attention as a key-enabling technology in 3D integration of integrated circuits. 3D TSVs allow heterogeneous assembly of multiple, disparate dies to form a stacked-die microsystem, and is an attractive way to empower "More than Moore" applications. It provides technical solutions to achieve better electrical performance and lower power consumption for integrated circuits [1, 2]. Copper is widely used as the via-filling material because copper is compatible with the back-end of line (BEOL) processes along with appropriate electrical and mechanical properties. However, due to the relatively large mismatch of coefficients of thermal expansion (CTE) between silicon (2.3 ppm/$^o$C) and copper (17 ppm/$^o$C) [3], embedded Cu TSVs in silicon ICs can induce a large thermo-mechanical stress during temperature ramps in fabrication. This stress has detrimental effects such as cracking, delamination and malfunctioning of individual transistors [4-8]. It can also vary the carrier mobility in devices through the piezoresistivity effect [9,10].

Recently, carbon nanotubes (CNTs) have emerged as an alternative TSV filling material. The advantages of CNTs lie in multiple aspects: (i) high current carrying capacity, (ii) low grain-boundaries, i.e., low internal-scattering resistance, (iii) high thermal-conductivity (1750-5800 W·mK$^{-1}$), which exceeds that of copper (400 W·mK$^{-1}$) by as much as 15 times, and is therefore ideal for dissipating heat from active hotspots to heat-sinks, and (iv) lower coefficient of thermal expansion (CTE) at ±0.4 ppm/$^o$C as compared to that of Cu (17 ppm/$^o$C) or Si (2.3ppm/ $^o$C). There has been some experimental and computational works using CNT as TSV filling material [11-14]. It was reported that even air-filled CNT bundles had conductivities a few times higher than that of Cu [15]. Due to the negative axial CTE of CNTs [11], it is expected that CNT should have different thermo-mechanical behaviors during the thermal processes compared to Cu, which may overcome the reliability problems such as via protrusion and interfacial delamination caused by Cu-filled TSVs. However, the stress field around CNT-filled TSVs has not been studied thoroughly.

Raman spectroscopy-based temperature dependent stress study was used in our recent work on Cu-filled TSVs with 5 to 10 μm in diameter [16, 17]. Together with stress simulations, the work provided evidences that the near surface stresses are not only due to the CTE-mismatch effect, but also from the pre-existing stress before the Cu filling [16]. However, in those works, the pre-annealed stress distributions or Cu-etched TSVs were not available for comparisons.

In this work, we fabricated much larger (40-μm-diameter) Cu filled TSVs. The 40-μm-diameter Cu TVSs were used instead of 5- or 10-μm-diameter Cu TVSs such that the Raman excitation laser spot size (0.8 μm) is much smaller than the TSV diameter. By doing this, the measured stress can be treated as stress at a point rather than an area averaged stress value, which is more accurate. The pre-annealing and post-annealing stress distributions were measured, and Cu-etched TSVs were made for comparisons to prove our hypothesis on the "pre-existing" stress in Ref. [16]. CTE-mismatch stress was calculated using a finite element analysis (FEA) tool. Moreover, CNT-filled TSVs and void vias without CNT filling were studied. The pre-existing stress and the CTE-mismatch stress of both types of TSVs were compared.

This work was funded by the Natural Science and Engineering Research Council of Canada (NSERC) and by the Agency for Science Technology and Research (A*STAR) through the public sector fund (#1121202004).

Ye Zhu, Yiheng Lin and Guangrui (Maggie) Xia are with the Department of Materials Engineering, the University of British Columbia, Vancouver, BC, Canada V6T1Z4 (e-mail: gxia@mail.ubc.ca). Kaushik Ghosh and Chuan Seng Tan are with the School of Electrical and Electronic Engineering, Nanyang Technological University, Singapore 639798 (e-mail: tancs@ntu.edu.sg). Hong Yu Li is with the Institute of Microelectronics, A*STAR, Singapore 138632.

## 2. Experiments

*2.1 Fabrication of Cu-filled TSVs and Cu-etched TSVs*

In this work, 40-μm-diameter Cu TSV arrays were fabricated on an 8-inch Si (100) wafer. 40-μm-diameter vias were etched to 60 μm depth on the Si substrate by a BOSCH process. The pitch varied from two to three times of the via diameter (i.e., from 80 to 120 μm). 1 μm thermal $SiO_2$ liner was grown at 1100°C for the isolation of TSV from the Si substrate after the via etch and wet clean. 100% TSV liner step coverage was achieved through the thermal oxidation. The tantalum (Ta) barrier layer of 20 nm and the Cu seed were sputtered and followed by Cu electroplating to solid fill the TSV. Chemical mechanical polishing (CMP) with high polish rate slurry was used to polish down the top oxide layer on Si surface and the top part of Cu. Fig. 1 (a) illustrates the schematic diagram of the cross-section of a 40-μm-diameter Cu TSV after the CMP step. The wafer was then thinned down to 100 μm from TSV backside by a back grinding system after TSV metallization. Cu annealing was performed at 300 ℃ for 60 minutes in nitrogen ambient. Fig. 1 (b) shows the optical microscopy plane view of the TSV arrays after the whole fabrication and annealing process.

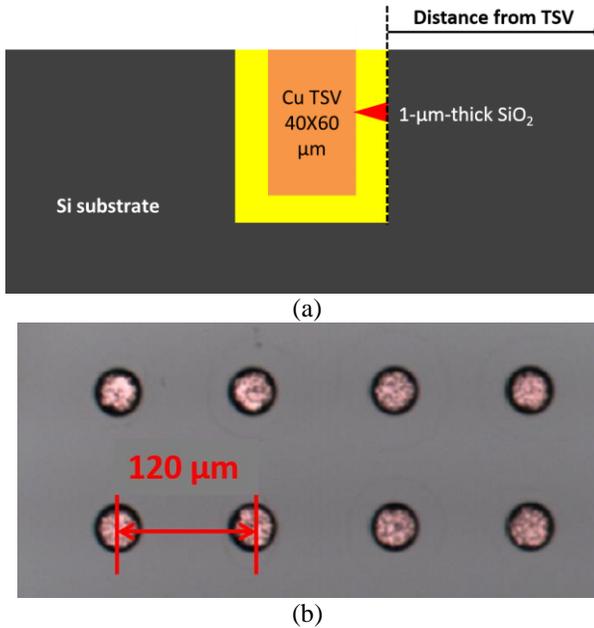

(a)

(b)

Fig. 1. (a) Schematic diagram of the cross-section of a 40-μm-diameter Cu TSV after the CMP and substrate thinning step, (b) optical microscopic plane view of the TSV arrays (pitch=120 μm) after the fabrication and annealing process (under a 10x objective). The distance from TSV is illustrated in (a), and the origin is at the $SiO_2$ layer and Si interface. As the Ta barrier layer is very thin (20 nm), we neglected this layer in all the figures.

For comparison, we fabricated the Cu-etched TSVs as control samples by etching the Cu filling after the 300 ℃ 60 min annealing. Ferric chloride ($FeCl_3$) solution was used in the Cu etching. At room temperature 23 ᵒC (RT), the annealed Cu-filled TSV sample was immersed in the $FeCl_3$ solution for 24 hours and then taken out for a surface clean by a diluted hydrogen chloride solution.

*2.2 Fabrication of CNT-filled TSVs and the Void Vias*

In this work, CNT-filled TSVs and void vias were fabricated in separated regions on 6-inch Si (100) wafers. The average diameter (*Φ*) of the vias is 12 μm with a fixed depth-to-diameter aspect ratio (AR=8). The average pitch is 24 μm. The fabrication steps of the void vias and liners before CNT filling are: (1) 500 nm thermal silicon dioxide growth on a bare Si wafer as a hard mask during the via etch step, (2) photolithography to pattern the photoresist for the via etch, (3) reactive ion etch of the silicon dioxide to open holes prior to the via etch, (4) deep reactive ion etch to form 85 micron deep void vias using BOSCH process, (5) remove photoresist, (6) remove the hard mask oxide layer by a wet etch step with HF solution, and (7) 500 nm thermal silicon dioxide growth at 1000 ℃ for 110 mins by wet oxidation to form the via liner.

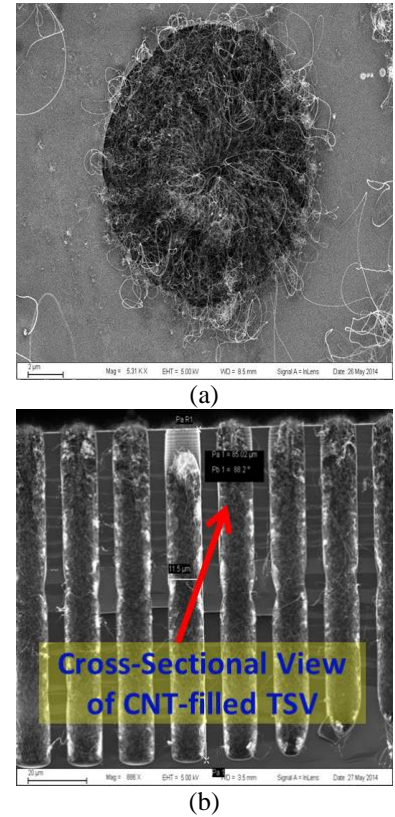

(a)

(b)

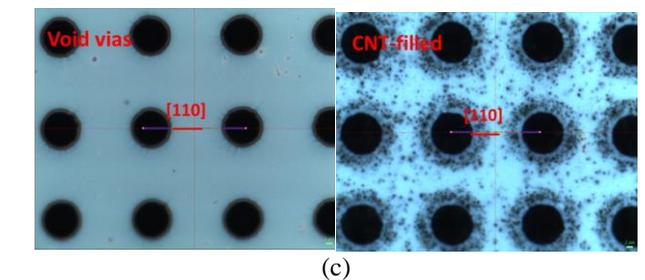

(c)

Fig. 2. (a) SEM top view of a CNT-filled TSV, (b) cross-sectional image of the CNT-filled TSVs, (c) optical microscopic plane view of void vias and CNT-filled TSVs after the fabrication process. The red solid lines indicate the Raman 1D scan ranges.

The catalyst used for the CNT growth was iron (Fe) nanoparticles similar as the approach reported in Ref. [13]. The steps for the catalyst formation are: (1) the Si substrate was merged in an $Fe_2(NO_3)_3$ solution with mild sonication allowing the solution to go inside the void vias, (2) the sample was dried under optimal heating in hydrogen environment to form Fe nanoparticles, and (3) the substrate was placed with the surface facing down touching a HF solution to remove the surface oxide and Fe layer without disturbing the Fe nanoparticles at the bottom of the vias. This is the reason that CNT only grew inside the vias, but not on the wafer surface. Only one half of the void vias saw the catalyst formation steps, and the other half void vias were used as the control group.

After that, high-density carbon nanotubes were grown inside one half of the void vias through two steps: (1) an annealing before CNT growth at 450 ℃ for 10 min in $H_2$ flow for CNT catalyst activation, and (2) the CNT growth at 550 ℃ for 30 min under $H_2/C_2H_2$ flow and cooling down to the RT. Most TSV holes were filled by carbon nanotubes completely as shown in Fig. 2 (a) and (b). We assumed that at the growth temperature 550 ℃, the newly grown CNTs are stress-free. Fig. 2 (b) shows a cross-sectional image of the CNT-filled TSVs and some process non-uniformity can be seen. Fig. 2 (c) shows the optical microscopy plane view of void vias and CNT-filled TSVs after the fabrication process.

*2.3 Stress Characterization with Raman Spectroscopy*

A LabRAM HR Raman con-focal microscope by Horiba Scientific was used for the Raman measurements. The input laser wavelength used was 442 nm, and the focused laser spot size is about 0.8 μm with a 100× objective. Four temperatures, 23, 50, 75 and 100 ℃, were used for all samples. One-dimensional (1D) Raman scanning was used to measure the stress distribution in Si. The scanning paths were along [110] directions as shown in Fig. 2 (c). For the Cu-etched and Cu-filled vias, the scanning distance and the step size were 30 μm and 1 μm, and for void vias and the CNT-filled vias, those were 10 μm and 0.25 μm respectively. Then, the sum of the radial stress and the circumferential stress $\sigma_r + \sigma_\theta$ was calculated from the Si Raman peak position. Details on the Raman measurements and stress calculations can be found in Ref. [16]. With careful measurements, the error bar of $\sigma_r + \sigma_\theta$ in our work is estimated to be ±15 MPa [16].

Simplified process flows for two kinds of TSVs with different filling materials (Cu and CNT) and their respective control samples are shown in Fig. 3. Stress measurement steps are shown in Fig. 3 as well.

### 3. Results

*3.1 Stress distribution around Cu-filled TSVs and Cu-etched TSVs*

The near-surface $\sigma_r + \sigma_\theta$ distributions in Si at the close vicinity of Cu-filled TSVs and Cu-etched TSVs from Raman measurements at RT are shown in Fig. 4 (a). It was observed that before the annealing process the near-surface stress in Si surrounding the Cu-filled TSV has a residual stress: a tensile stress peak at about 3 μm away from the Si/TSV liner interface. This tensile peak dropped after 300 ℃ annealing, and returned to the pre-annealing value after Cu etching. The stress curves for Cu-filled TSVs before annealing and TSVs after Cu etching are very close. It gave strong evidence that the residual stress in the Cu-filled TSV structures before annealing (which is just after the Cu plating step) is not related to Cu. It should have been induced before the Cu plating. We suspect that this stress is originated from the thermal growth of the oxide liner.

Fig. 4 (b) and (c) show the temperature-dependent near-surface Si $\sigma_r + \sigma_\theta$ distributions for annealed Cu-filled TSVs and Cu-etched TSVs, respectively, at four temperatures, 23, 50, 75 and 100 ℃. It can be observed that CTE-mismatch effect dominates the compressive regime of the stress distribution in Si surrounding the annealed Cu-filled TSVs. However, within experimental errors, the stress distribution in Si around the Cu-etched TSVs did not change with temperature, indicating that the pre-existing residual stress is constant in this temperature range. It was not possible to measure the pre-annealed Cu-filled TSV stress in different

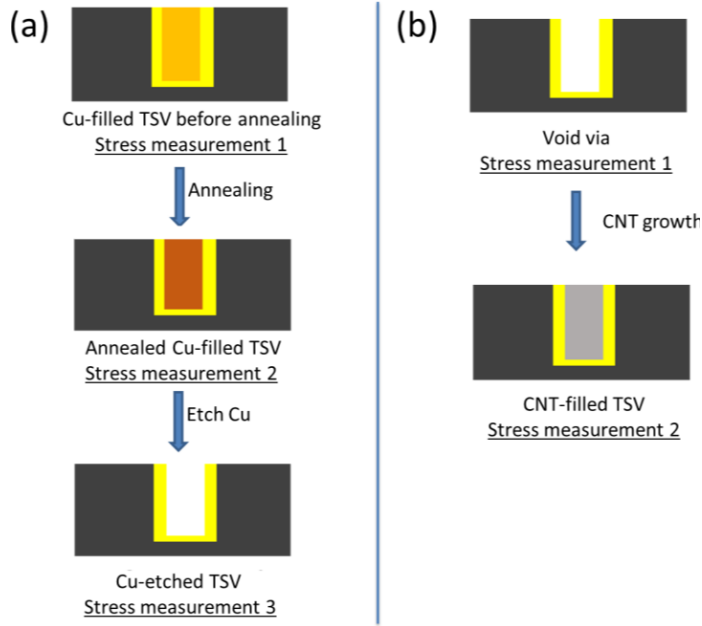

Fig. 3. Simplified process and stress measurement flows for two kinds of TSVs with different filling materials. (a) Cu-filled TSVs, and (b) CNT-filled TSVs.

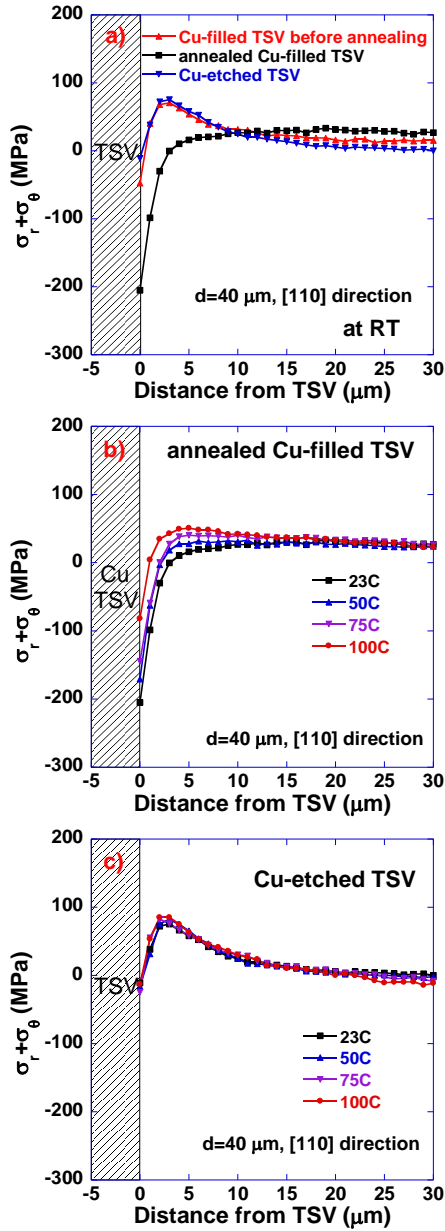

Fig. 4. $\sigma_r + \sigma_\theta$ distributions as a function of distance from the 40-μm-diameter TSV edge along the [110] direction: (a) for Cu TSV before annealing, after annealing and Cu etching, respectively, at RT; (b) for annealed Cu-filled TSV, (c) for Cu-etched TSV at four different temperatures (23, 50, 75 and 100 °C).

elevated temperatures as it introduces the annealing effects such as Cu grain growth.

*3.2 CTE-mismatch Stress Simulations*

The stress distributions in Si around Cu-filled TSVs were simulated by a standard 3-D FEA tool called COMSOL$^{TM}$. The details on the stress simulations can be found in Ref. [16]. The Ta barrier layer is 20 nm thick, which is too thin to have any significant impact on the stress distribution, and thus was neglected in the stress simulations.

Fig. 5 (a) shows the experiment results (Curve 1, same data as in Fig. 4 (a)) in comparison with the simulation results of $\sigma_r + \sigma_\theta$ (Curve 2) at the laser penetration depth (0.2 μm) for a Cu-filled TSV along [110] direction at RT. Compared with Curve 1 under the same temperature, Curve 2 present a disagreement in the region 1 μm < x < 10 μm, where Curve 1 show a tensile stress peak at ~3 μm away from the Si/TSV liner interface. The reason we proposed for this tensile peak is a "pre-existing" stress which has been induced in Si around the TSVs before Cu filling. Fig. 5 (b) shows the differential values between Curve 1 and Curve 2 in comparison with a Cu-filled TSV before annealing (Curve 3) with the data from Fig. 4 (a) as well. From this figure, we can see that they agree with each other within the experimental errors. This agreement verifies the hypothesis we proposed that the total stress is the sum of the "pre-existing" stress and the CTE-mismatch stress.

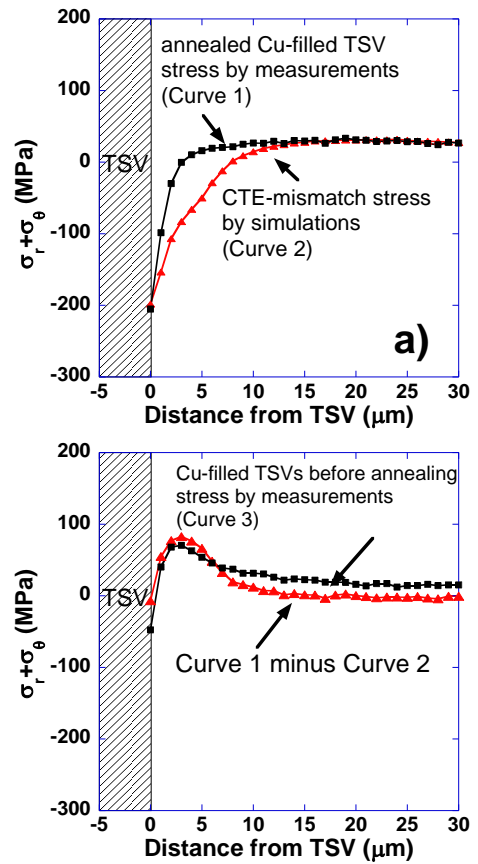

Fig. 5. (a) The simulated CTE-mismatch stress $\sigma_r + \sigma_\theta$ (Curve 2) in comparison with the experiment results (Curve 1) at the laser penetration depth (0.2 μm) for a 40-μm-diameter Cu-filled TSV along [110] direction at 23 °C; (b) Curve 1 minus Curve 2 in comparison with the measured stress for a Cu-filled TSV before annealing (Curve 3).

*3.3 Stress distribution around CNT-filled TSVs and void vias*

Fig. 6 (a) shows the comparison of stress distributions between the void vias and the CNT-filled samples at RT. The difference in the stress curves is within measurement error bar, indicating the pre-existing residual stress is dominant in the CNT-filled sample. These measurements prove our hypothesis that the processing history before via filling can induce a pre-existing stress in Si around the TSVs.

Both void vias and CNT-filled vias have seen the 450 ℃ catalyst activation and 550 ℃ CNT growth steps. During the CNT growth, the newly grown CNT should be stress free. After the cooling step, the Si stress is almost back to the pre-existing stress, which means that the catalyst activation and the CNT growth and cooling step didn't introduce any additional stresses. There are two possible explanations for this phenomenon: (1) the CTE of CNT filling is close to that of Si; (2) there is a weakly bounded interface or air gaps between the CNT filling and the TSV liner, which means that when the CNTs shrank during cooling, they did not drag the surrounding Si, consistent with the SEM image in Fig. 2 (b).

Next, we heated the samples up, and the near-surface Si $\sigma_r + \sigma_\theta$ distributions at the elevated temperatures are shown as Fig. 6 (b) and (c), respectively. The difference for different temperatures is (~15 MPa) within the error bar of the measurements. Fig. 6 (c) shows that within experimental errors, the stress distribution of the CNT-filled sample did not change much with the temperature increase, indicating that the CTE-mismatch stress induced by the CNT-filled TSV is negligible. This observation is consistent with the explanation in (1) and (2) as well.

## 4. Discussions

Fig. 7 shows the comparison of pre-existing stresses for the CNT-filled TSVs, 40-μm-diameter Cu-filled TSVs and 10-μm-diameter Cu-filled TSVs (data from Ref. [16]) at room temperature. From this comparison, we can see that the pre-existing stresses are different for different TSV samples, which means that it is processing dependent. The 40-μm-diameter Cu-filled TSV sample has a tensile pre-existing stress larger than 10-μm-diameter Cu-filled TSV sample, whose liner was deposited by plasma-enhanced chemical vapor deposition. The positions of the tensile peaks of pre-existing stresses for the CNT-filled TSVs, 10-μm-diameter and the 40-μm-diameter Cu-filled TSVs are all around 2 μm, and don't scale with the their diameters. On the contrary, the simulated CTE-mismatch stresses for two batches of Cu-filled TSVs scale with their diameters. This is expected, as in the CTE-mismatch stress simulation setup, one TSV in a large Si substrate was simulated, and therefore the stress distribution scales with the TSV diameter. CTE-mismatch stresses for two batches of Cu-filled TSVs are slightly different due to their structure details. The 10-μm-diameter Cu-filled TSVs have a top SiO$_2$ layer on the Si wafer surface, while for the 40-μm-diameter Cu-filled TSVs, the surface oxide was removed.

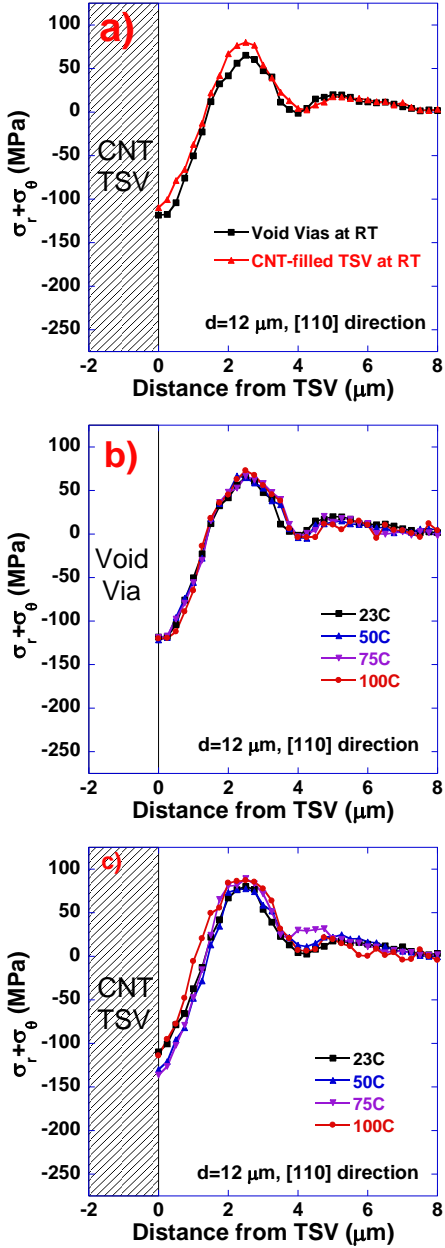

Fig. 6. $\sigma_r + \sigma_\theta$ distributions as a function of distance: (a) the stress comparison between the void vias and the CNT-filled TSVs at RT; (b) for the void vias, and (c) for the CNT-filled TSVs at four different temperatures (23, 50, 75 and 100 ℃).

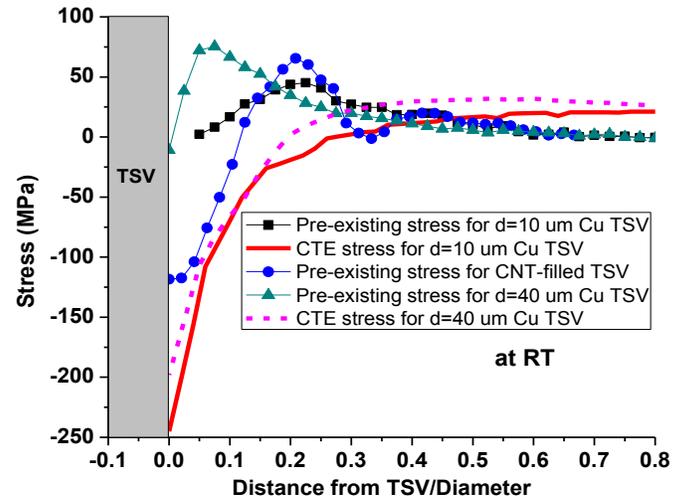

Fig. 7. The comparison of pre-existing stresses for the 12-μm-diameter CNT-filled TSV, 10-μm-diameter Cu-filled TSV and 40-μm-diameter Cu-filled TSV at room temperature.

Previous study suggested that with liners made from low-k dielectric materials, some densification may happen which impacts the TSV stress [18]. For thermally grown SiO$_2$ liners, however, we don't expect any densification at the annealing temperature studied. It is well-known that thermal oxidation can introduce stress in the oxide and the nearby Si [19], and we suspect that the pre-existing stress is relevant to the thermal SiO$_2$ growth of the via liners for the 40-μm-diameter Cu-filled TSV and CTN-filled TSVs. Due to the cylindrical and wavy sidewall shape and the difficulty in three-dimensional stress simulations during thermal oxidation, we will leave this topic for future study.

## 5. Conclusions

In summary, micro-Raman spectroscopy was employed to study the near-surface stress distributions and origins around Cu-filled and CNT-filled TSVs at elevated temperatures. Our results proved that the stresses near TSVs are originated from two sources: 1) pre-existing stress before via filling, and 2) CTE-mismatch-induced stress. CTE-mismatch-induced stress is shown to dominate the compressive regime of the near-surface stress distribution around Cu-filled TSV structures, while pre-existing stress dominates the full range of the stress distribution in the CNT-filled TSV structures. We recommend using a liner growth technique that introduces less stress in Si. Once the pre-existing stress is minimized, the total stress around CNT-filled TSVs can be minimized accordingly. Therefore, compared to Cu-filled TSVs, CNT-filled TSVs hold the potential to circumvent the hassle of stress-aware circuit layout and to solve the stress-related reliability issues.